\documentclass[10pt,conference]{IEEEtran}
\IEEEoverridecommandlockouts

\usepackage{cite}
\usepackage{amsmath,amssymb,amsfonts,bm}
\usepackage{algorithmic}
\usepackage{graphicx}
\usepackage{textcomp}
\usepackage{xcolor}
\usepackage{booktabs}
\usepackage{listings}
\usepackage{subfigure}
\usepackage{caption}
\usepackage{balance}

\usepackage{hyperref}
\hypersetup{
    colorlinks=true,
    linkcolor=blue,
    filecolor=blue,      
    urlcolor=blue,
    citecolor=blue
}
\definecolor{codegreen}{rgb}{0,0.6,0}
\definecolor{codegray}{rgb}{0.5,0.5,0.5}
\definecolor{codepurple}{rgb}{0.58,0,0.82}
\definecolor{backcolour}{rgb}{0.95,0.95,0.92}
 
\lstdefinestyle{mystyle}{
    commentstyle=\color{codegreen},
    keywordstyle=\color{purple},
    numberstyle=\tiny\color{codegray},
    stringstyle=\color{violet},
    basicstyle=\ttfamily\footnotesize,
    breakatwhitespace=false,         
    breaklines=true,                 
    captionpos=b,                    
    keepspaces=true,                 
    showspaces=false,                
    showstringspaces=false,
    showtabs=false,                  
    tabsize=2,
    frame=single,
    frameround=fttt,
    xleftmargin=5pt,
    xrightmargin=5pt,
    showstringspaces=false
}
 
\usepackage{fancyhdr}
\usepackage{multirow}

\begin{document}

\title{CORE: Automating Review Recommendation for Code Changes
\thanks{\IEEEauthorrefmark{1} Cuiyun Gao is the corresponding author.}
}

\author{
\IEEEauthorblockN{Jing Kai Siow, Cuiyun Gao\IEEEauthorrefmark{1}, Lingling Fan, Sen Chen, Yang Liu}
\IEEEauthorblockA{
   School of Computer Science and Engineering, Nanyang Technology University, Singapore
   \\
   jingkai001@e.ntu.edu.sg,
\{cuiyun.gao,yangliu\}@ntu.edu.sg, \{ecnujanefan,ecnuchensen\}@gmail.com
}
}

\maketitle

\begin{abstract}
Code review is a common process that is used by developers, in which a reviewer provides useful comments or points out defects in the submitted source code changes via pull request. Code review has been widely used for both industry and open-source projects due to its capacity in early defect identification, project maintenance, and code improvement. With rapid updates on project developments, code review becomes a non-trivial and labor-intensive task for reviewers.
Thus, an automated code review engine can be beneficial and useful for project development in practice. 
Although there exist prior studies on automating the code review process by adopting static analysis tools or deep learning techniques, they often require external sources such as partial or full source code for accurate review suggestion. In this paper, we aim at automating the code review process only based on code changes and the corresponding reviews but with better performance. 

The hinge of accurate code review suggestion is to learn good representations for both code changes and reviews. To achieve this with limited source, we design a multi-level embedding (i.e., word embedding and character embedding) approachto represent the semantics provided by code changes and reviews.The embeddings are then well trained through a proposed attentional deep learning model{, as a whole named CORE}. We evaluate the effectiveness of CORE on code changes and reviews collected from 19 popular Java projects hosted on Github. Experimental results show that our model CORE can achieve significantly better performance than the state-of-the-art model (DeepMem), with an increase of 131.03\% in terms of Recall@10 and 150.69\% in terms of Mean Reciprocal Rank. Qualitative general word analysis among project developers also demonstrates the performance of CORE in automating code review.
\end{abstract}

\graphicspath{ {./fig/} }

\section{Introduction}
Code review is a process that involves manual inspection of source code revisions, either by peers or by colleagues, before pushing the revisions to the live system. This process is commonly used by software engineers and open-source project developers to ensure that the new code revisions  
 
{do not introduce errors}
and adhere to the guidelines of the projects. A formal and structured framework, commonly known as Fagan Inspection~\cite{5388086} proposed in 1976, to provide developers a way to identify defects in development phase. Code review aids developers in avoiding errors and finding defects during the development or maintenance phase. However, modern code review has evolved from finding bugs to providing maintainability and understanding of source code and their changes \cite{47025,Beller:2014:MCR:2597073.2597082}. 

Although finding defects and errors are still one of the priorities~\cite{fan2018large,fan2018efficiently}, developers from industries value more on motivations, such as transferring of knowledge, understanding of source code and code improvement~\cite{47025,bird2013expectations,greiler2016code}. Several aspects on code reviews have been explored by software engineering, such as link graph analysis~\cite{Hirao:2019:RLG:3338906.3338949}, sentiment analysis~\cite{8115623}, decision making \cite{Shi2019AutomaticCR,8104731}, reviewer recommendation~\cite{Hannebauer:2016:ARC:2970276.2970306,Asthana:2019:WAR:3338906.3340449,6606642}, matching identical code review~\cite{Gupta2018IntelligentCR}, security bug analysis~\cite{icse2019, chen2018mobile,ausera2018}, and motivations and objectives of code reviews~\cite{47025,bird2013expectations}.

{As the projects become more sophisticated over time,}
the amount of code change increases daily. Code reviewers often have difficulty in allocating a huge amount of time in performing code review. Furthermore, to conduct code review, reviewers need to understand the purpose and history of the code before they can perform a fair evaluation of the code changes~\cite{bird2013expectations}. This results in a huge amount of time needed for the code review process. Many research studies \cite{Hannebauer:2016:ARC:2970276.2970306, 6606642, Gupta2018IntelligentCR} aim to reduce the workload of reviewers such as recommending the best reviewer or even automating the code review generation process. One specific work that is related to our paper, by Gupta and Sundaresan \cite{Gupta2018IntelligentCR}, uses basic LSTM models to learn the relation between code changes and reviews.

In practice, developers often use code collaboration tools, such as Gerrit\cite{Gerrit} and Review Board\cite{ReviewBoard}, to assist them in the process of code review. Their functionalities often include showing changed files, allowing reviewers to reject or accept changes and searching the codebase. Some research work, \textit{e.g.,} ReviewBot\cite{6606642}, incorporates static analysis tools to publish reviews automatically. These static analyzers detect defect code and unconventional naming by the submitter and publish them as part of code review. Static analyzers require a comprehensive set of rules that allow them to detect defective source code. In comparison with human generated reviews, reviews by static analyzers are more rigid and have difficulty in finding errors that are emerged outside of their heuristic rules\cite{static}. Other work, like DeepMem\cite{Gupta2018IntelligentCR}, aims to recommend reviews by learning the relevancy between source code and review. However, these methods often require a large amount of additional sources, such as full or partial source code. This additional information might not be available at all times, for instance, submitting pull request or commits.

In this paper, we propose a novel deep learning model for recommending relevant reviews given a code change. We name the whole \underline{CO}de \underline{RE}view engine as CORE. CORE is built upon only code changes and reviews without external resources. 
 
{Our motivation is to reduce the workload of developers by providing review recommendation without human intervene. By automating the code review process, developers can correct their code as soon as possible, hence, reducing the time between each revision of code changes.}
The challenging point of automating code review is to learn good semantic representations for both sources. Although word embeddings\cite{Mikolov:2013:DRW:2999792.2999959, word2vec-2} have been proven useful in representing the semantics of words, they may fail in capturing enough semantics of code since out-of-vocabulary words are often introduced into the project along development phase\cite{DBLP:journals/corr/AllamanisPS16}. To overcome this challenge, we propose a two-level embedding method which combines both word-level embedding and character-level embedding \cite{DBLP:journals/corr/ZhangL15} for representation learning.

We then predict the consistency (\textit{i.e.}, relevancy score) between two sources with an attentional neural network, where the attention mechanism~\cite{Bahdanau2014NeuralMT} learns to focus on the important parts of the two sources during prediction. Experimental analysis based on 19 popular Java projects hosted on GitHub demonstrates the effectiveness of the proposed CORE. CORE can significantly outperform the state-of-the-art model by 131.03\% in Recall@10 and 150.69\% in Mean Reciprocal Rank. 

Qualitative analysis also shows that project developers in the industry are interested in our work and agreed that our work are effective for practical software development.

In summary, our contributions are as follows:
\begin{itemize}
  \item We propose a novel deep learning model, CORE, for recommending reviews given code changes. CORE combines multi-level embedding, \textit{i.e.,} character-level and word-level embedding, to effectively learn the relevancy between source code and reviews. CORE can well learn the representations of code changes and reviews without external resources.
  
  \item Extensive experiments demonstrate the effectiveness of CORE against the state-of-the-art model. We also evaluate the impact of different modules in our model, from which multi-level embedding is proven to be more effective in improving the performance of CORE.

  \item We build and provide a benchmark dataset containing 57K pairs of $<$code change, review$>$
  
  collected from 19 popular Java projects hosted on GitHub. We release our dataset on our website\footnote{\url{https://sites.google.com/view/core2019/}} for follow-up code review-related tasks.
  
\end{itemize}

\section{Preliminary} \label{sec:preliminary}

\subsection{Code Review}
Traditionally, reviews for source code and code revisions are contributed by humans manually. These reviews are generally in natural languages and short sentences. More often than not, these reviews serve as suggestions or comments for the authors, informing them if there are any errors or concerns with the newly submitted code. There are several tools in the market, such as Gerrit~\cite{Gerrit}, Review Board~\cite{ReviewBoard} and Google's Critique\cite{47025} that could help the developers in facilitating better reviewing process and providing additional reviewing tools.

Both industry and open-source projects adopted code review process to improve their code quality and reduce the number of errors to be introduced into the codebase. In this paper, we explore the code reviews in open-source projects, for instance, projects that are hosted on GitHub. GitHub allows others to submit their changes to the project through a function known as Pull Request~\cite{PullRequest}. Meanwhile, the review would be required for these changes to determine whether they should be merged into the main branch of the project.
The author of the pull request will request a review for the submitted code changes from the main contributors of the projects. These main contributors are usually experienced developers that involved deeply in the the requested open-source project. 

\subsection{Motivating Examples}

Listing \ref{fig-review-1} shows an example of a code review that we aim to match. As we can see, given a code change snippet, the reviewer suggested that instead of using \textit{completions.add(completion)}, the author should use \textit{java.util.Collections.addAll()}. This suggestion is useful in keeping the codebase consistent and could help to reduce unnecessary mistakes. Automating such reviews could allow reviewers to save their precious time and efforts.

Another example in Listing \ref{fig-review-2} shows that a better naming convention should be used instead of \textit{tcpMd5Sig()}. Such trivial reviews might be very simple but could be very time-consuming for reviewers. Hence, we aim to reduce the workload for reviewers by finding such useful and practical reviews that are commonly used throughout the open-source projects. 

\begin{lstlisting}[language=Java,caption={Review regarding API replacement},label={fig-review-1}]
// Code changes
+     completion.add(completion);
+    }
+  }
+  for (OffsetCommitCompletion completion: completions){
------------------------------------------------
// Review
    "Consider using java.util.Collection.addAll()"
\end{lstlisting}

\begin{lstlisting}[language=Java,caption={Review regarding naming convention},label={fig-review-2}]
// Code changes
+  private volatile Set<InetAddress> tcpMd5Sig = Collections.emptySet();
------------------------------------------------
// Review
    "Could this field and the tcpMd5Sig() method have a better name? It does not contain any signature but a set of addresses only."
\end{lstlisting}

\subsection{Word Embedding}

Word embedding are techniques on learning how to represent a single word using vector representation. Each word is mapped to a unique numerical vector. For instance, given a word ``\textit{simple}", we embed it into a vector of $\bm{x_1} = [0.123, 0.45, ..., 0.415]$ where $\bm{x_1}$ is a vector of length $n$. Commonly, words that have similar meanings tend to have lower distance in the embedded latent space.
There are several techniques that employ deep learning and deep neural network to learn richer word representations, such as Word2Vec \cite{Mikolov:2013:DRW:2999792.2999959} and ELMo \cite{DBLP:journals/corr/abs-1802-05365}. Word2Vec uses fully connected layer(s) to learn the context around each word and outputs a vector for each word, while ELMo uses Bi-LSTM to learn deeper word representations. Word embedding are often pre-trained to ensure that downstream tasks can be performed as efficiently as possible.

\subsection{Long Short-Term Memory}
Long Short-Term Memory (LSTM) is a type of recurrent neural network that is typically used to learn the long term dependencies in sequence data, such as time-series or natural language. It is commonly used in natural language processing as it learns the dependencies in a long sequence of words. Assume that the embedding sequence is in form of $X = [x_1, x_2, ..., x_2]$ where $X$ represents the embedded sequence of tokens, LSTM computes the current output $o_t$ based on its previous state and the memory cell state.
\begin{equation}
\begin{split}
o_t = tanh({h_{t-1}, x_t})
\end{split}
\end{equation}
 
where $h_{t-1}$ represents the previous state and $x_t$ represents the current input at time-step $t$. The $\tanh(\cdot)$ is a typical activation function for learning a non-linear adaption of the input. These outputs from LSTMs contain latent representations of the tokens in the sentence. 

\subsection{Attention Mechanism}
\begin{figure}
	\centering
	\includegraphics[width=0.4\textwidth]{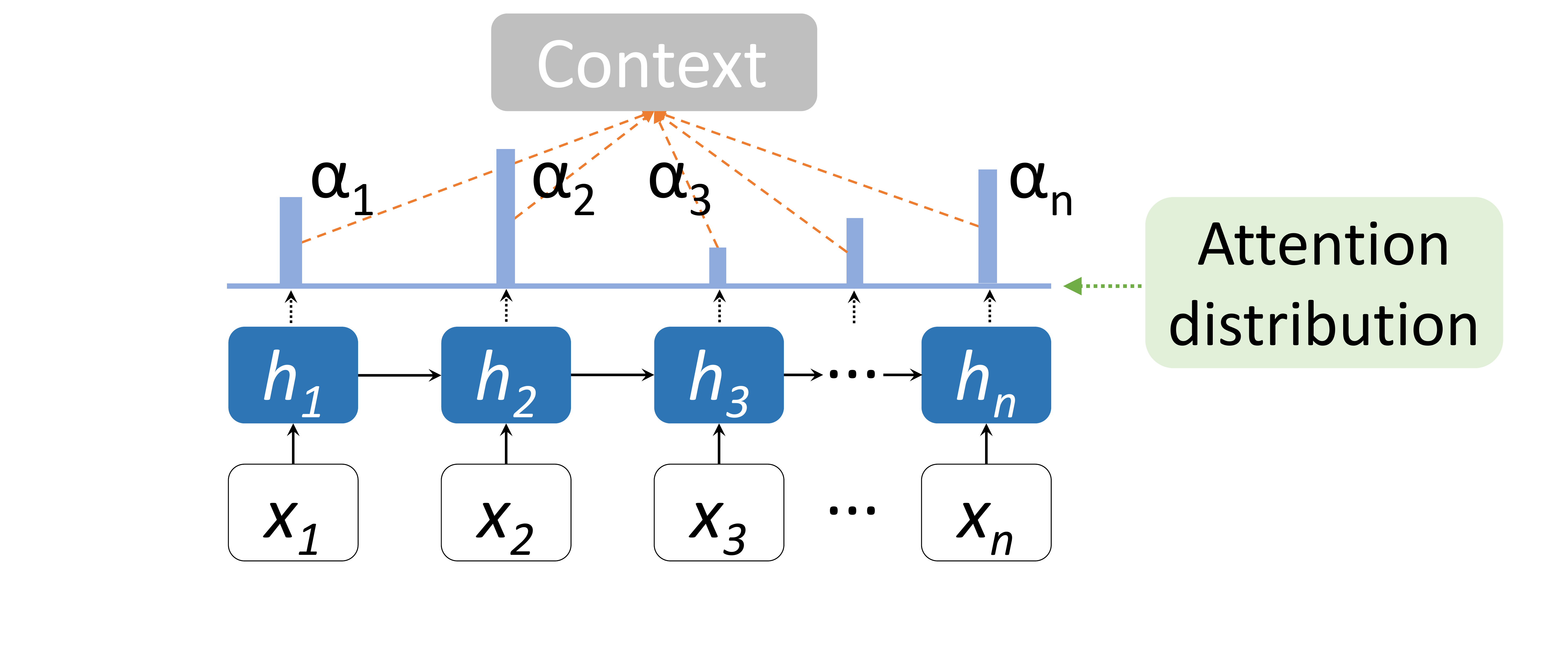}
	\caption{Attention mechanism}
	\label{fig-attention}
\end{figure}

Attention mechanism is proposed by Bahdanau\cite{Bahdanau2014NeuralMT} and it is first used in natural language processing, mainly in neural machine translation. Attention mechanism computes a context vector for each sentence. It allows us to have better representation and provides a global context for each sentence that is beneficial to the relevancy learning task in our work. It is computed based on the outputs of LSTMs and learned weights, hence, giving it higher-level representation on the whole sequence. The context vector, shown in Fig.~\ref{fig-attention}, is a weighted sum of the output of LSTMs by using attention weights, $\alpha_{ij}$.
\begin{equation}
\begin{split}
\alpha_{ij} = \frac{\exp{(e_{ij})}}{\sum_{k=1}^n {\exp{(e_{ik})}}}
\end{split}
\end{equation}

Equation 2 shows the formula for computing the attention weights, where $n$ is the total number of LSTM outputs. The alignment score, $e_{ij}$, goes through a softmax function to achieve the attention weights. This allows the mechanism to have a higher weight for variables that contribute more towards the success of the model. $e_{ij}$ can be learned using an activation function and learned weights.
\begin{equation}
\begin{split}
e_{ij} = \tanh(W_s S_t + b) \\
\end{split}
\end{equation}

where $S_t$ is the output of LSTM, $b$ and $W_s$ is the bias and weight. Finally, the context vector can be computed by using a sum of the attention weights and the alignment score.
\begin{equation}
\begin{split}
\mathbf{a}_t = \sum_{i}^{n} S_t \alpha_{ij} 
\end{split}
\end{equation}

\subsection{Problem Definition}
In this work, we define the comments or suggestions by the reviewers as ``\textit{reviews}'' (denoted by $\mathbf{R}$) and the submitted code as ``\textit{code changes}'' (denoted by $\mathbf{C}$). 
Given a snippet of code changes, we aim to find the top-k most relevant reviews in our dataset.
We model it as a recommendation/ranking problem such that each code snippet will have a list of reviews that are sorted according to their relevancy score. For each \textless $c_i, r_i$\textgreater $\in$  \textless $\mathbf{C},\mathbf{R}$\textgreater, a score ({\textit{i.e.}, $Rel(c_i,r_i)$}) that indicates the relevancy of the code changes $c_i$ and review $r_i$ will be learned through our proposed deep learning model, which is computed as follows:  
\begin{equation}
Rel(c_i,r_i) = F(c_i, r_i, \theta)
\end{equation}
where $\theta$ represents the model parameters that will be learned and improved in the training process.

\section{Approach} \label{sec:approach}
In this section, we present the overview of our proposed model, CORE, and design details that extend the basic attentional Long Short-Term Memory model for code review recommendation. We regard code change and corresponding review text as two source sequences and the relevancy score as the target. Fig.~\ref{fig-overview} shows the workflow of CORE.

\subsection{Overview}
\begin{figure*}[t]
		\centering
		\subfigure[Overall architecture of CORE]
		{\label{fig:a}\includegraphics[scale=0.6]{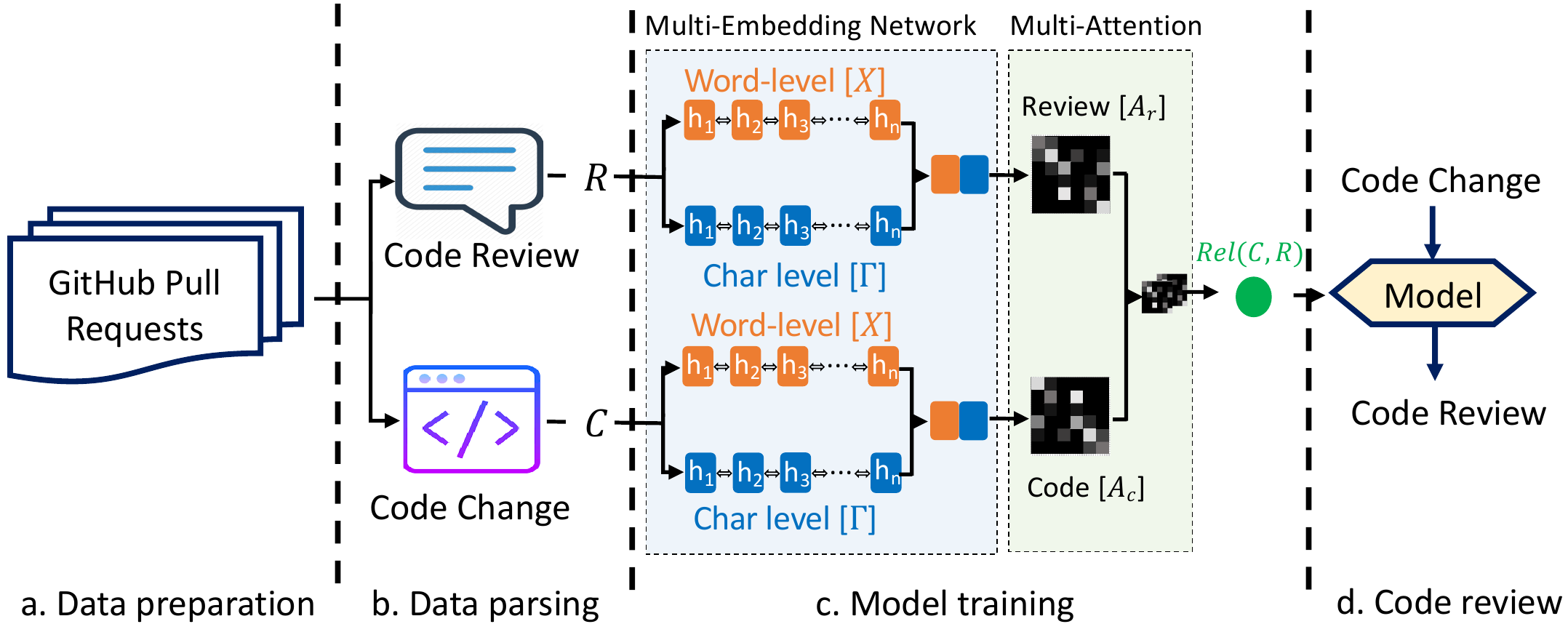}}
        \hspace*{\fill}
		\subfigure[Detailed structure of CORE]
		{\label{fig:b}\includegraphics[scale=0.6]{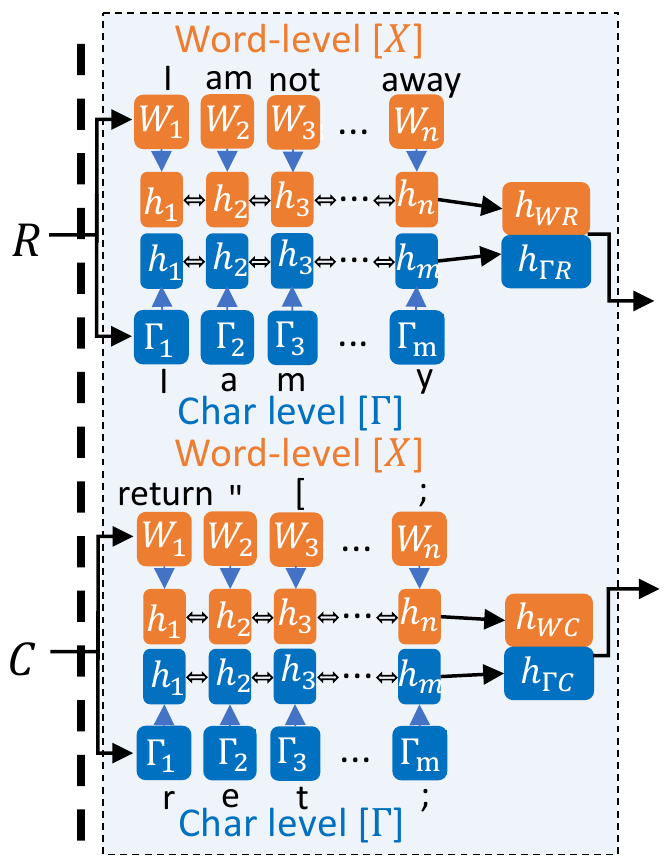}}
	\caption{Structure of the code review model.}
	\vspace{-4mm}
	\label{fig-overview}
\end{figure*}

As can be seen in Fig.~\ref{fig-overview}, the workflow of CORE mainly contains four steps, including data preparation, data parsing, model training, and code review. We first collect pull requests from GitHub and conduct preprocessing. The preprocessed data are parsed into a parallel corpus of code changes and their corresponding reviews, \textit{i.e.}, \textless $\bm{C},\bm{R}$\textgreater. Based on the parallel corpus of code changes and reviews, we build and train a neural-based {learning-to-rank} model. 
The major challenge during the training process lies in using limited information to well represent the two sources. Finally, we deploy our model for automated code review. In the following, we will introduce the details of the CORE model and the approach we propose to resolve the challenge.

\subsection{Multi-Level Embeddings}
To better represent code changes and review texts, we propose to combine the two levels of embeddings: \textbf{Word-level embedding} and \textbf{Character-level embedding}.

\subsubsection{Word-Level Embedding} 
Word embeddings are widely adopted in representing the semantics of tokens through training on a large text corpus. In this work, we adopt word2vec\cite{Mikolov:2013:DRW:2999792.2999959}, a distributed representation of words, as pre-trained embeddings for the words in code changes and reviews, and retrain them in the respective source. The detailed retraining processes are described in the following.

\noindent \textbf{Code review.} Code reviews are generally mingled with texts and project-specific tokens. The project-specific tokens usually appear only a few times, including function names, variable names, version numbers of the projects, and hash IDs of commits, so the embeddings of such tokens could not be well learnt and may bring noise into the ultimate review text representation. To alleviate such noise, we first tokenize the reviews and then replace the project-specific terms with placeholders. Specifically, we convert hash IDs with ``\textless HASHID\textgreater'', numerical digits with ``\textless NUM\textgreater'', version numbers with ``\textless VERSIONNUM\textgreater'', and URLs with ``\textless URL\textgreater''. The preprocessed reviews are employed to retrain review-specific word embeddings.

\noindent \textbf{Code change.} Code representation learning is a challenging task in natural language processing field because code usually contains project-specific and rarely-appearing tokens. In spite of their low frequency, they can also help understand the semantics of the code changes. For example, a variable name, ``SharedSparkSession", allows us to understand that the variable is closely related to a session and they are shared among multiple users. We adopt a typical parser, \textit{pygments}~\cite{Pygments}, to parse code changes into tokens. For example, given a source code statement in Java, ``\textit{private final int shuffleId;}", we parse them using pygments and append each token to a list, such that we will receive a list of token, ``[``private", ``final", ``int", ``shuffleId", ``;"]". The preprocessed code changes are then fed into the word2vec trainer, from which we can obtain code-specific word embeddings.

\subsubsection{Character-Level Embedding} The Out-of-Vocabulary (OOV) issue is 
common in code-related tasks~\cite{DBLP:journals/corr/abs-1711-09573, ijcai2019-552, Svyatkovskiy:2019:PAC:3292500.3330699, Hu:2018:DCC:3196321.3196334}. This is because code contains not only typical API methods but also randomly-named tokens such as variable names and class names. Although word-level embeddings could represent the semantics of the tokens in source code, the OOV issue still exists since low-frequency words are not included in the pre-trained word2vec model. To alleviate the OOV issue, we propose to combine character-level embeddings. The character-level embeddings are independent of tokens in the collection and represent the semantics of each character. In this work, we embed each character into a one-hot vector. For example, given a review, ``please fix this", we separate the sentence into a list of characters, [``p'', ``l'', ``e'', ``s'', ..., ``t'', ``h", ``i", ``s"]. We embed this list of characters using one-hot embedding, such that each character has their unique one-hot representation. 

\subsection{Multi-Embedding Network}
The multi-embedding network aims at jointly encoding the word-level and character-level representations for both code changes $\mathbf{C}$ and reviews $\mathbf{R}$, with details illustrated in Fig.\ref{fig:b}. We denote the two-level embeddings for both sources, denoted as $\mathbf{W}_C$ and $\mathbf{W}_R$ for the respective word-level embeddings, and $\mathbf{\Gamma}_C$ and $\mathbf{\Gamma}_R$ for the respective character-level embeddings. We adopt Bi-directional LSTMs (Bi-LSTMs) \cite{bilstm} to learn the sequence representations for word-level and character-level code changes, denoted as $\mathbf{h}_{WC}$ and $\mathbf{h}_{\Gamma C}$ respectively. The $\mathbf{h}_{WC}$ and $\mathbf{h}_{\Gamma C}$ are the last hidden states produced by the corresponding Bi-LSTMs. The final embeddings for code changes are defined as:

\begin{equation}
    \mathbf{h}_C = \tanh(\mathbf{W}^C[\mathbf{h}_{WC};\mathbf{h}_{\Gamma C}]),
\end{equation}

\noindent where $[\textbf{h}_{WC};\textbf{h}_{\Gamma C}]$ is the concatenation of the two-level representations, $\mathbf{W}^C$ is the matrix of trainable parameters, $C$ is the number of hidden units, and $\tanh(\cdot)$ is used as the activation function.

Similarly, we can obtain the two-level sequence representations $\mathbf{h}_{WR}$ and $\mathbf{h}_{\Gamma R}$ for code reviews $\mathbf{R}$. The final representations for reviews can be calculated as:

\begin{equation}
    \mathbf{h}_R = \tanh(\mathbf{W}^R[\mathbf{h}_{WR};\mathbf{h}_{\Gamma R}]),
\end{equation}

\noindent where $[\mathbf{h}_{WR};\mathbf{h}_{\Gamma R}]$ is the direct concatenation of the word-level and character-level embeddings for reviews $R$, $\mathbf{W}^R$ is the matrix of trainable parameters, and $R$ is the number of hidden units. For simplicity, we assume that the dimensions of the two-level embeddings, \textit{i.e.}, $\mathbf{h}_{WR}, \mathbf{h}_{\Gamma R}, \mathbf{h}_{WC}$, and $\mathbf{h}_{\Gamma C}$, are the same.

\subsection{Multi-Attention Mechanism}
To alleviate the influence of noisy input, we employ the attention mechanism \cite{Bahdanau2014NeuralMT} on the learned representations of code changes and reviews, \textit{i.e.}, $\mathbf{h}_C$ and $\mathbf{h}_R$. The attention mechanism can make the training process pay attention to the words and characters that are representative of code changes and reviews. As can be seen in Fig.~\ref{fig-overview}, both representations are further enhanced through attention layers:

\begin{equation}
    \mathbf{a}_{C} = \sum_{i}^{n} \mathbf{H}_C \alpha_{ij}^c
\end{equation}
\begin{equation}
    \mathbf{a}_{R} = \sum_{i}^{n} \mathbf{H}_R \alpha_{ij}^r 
\end{equation}

\noindent where outputs, \textit{i.e.}, $\mathbf{a}_{C}$ and $\mathbf{a}_{R}$, indicate the learned attended representations of code reviews and reviews respectively. The two attended vectors are then concatenated into a \textit{multi-attention} vector, $\mathbf{a}_{\mathbf{C},\mathbf{R}}$, which is finally trained for predicting relevancy scores $Rel(\mathbf{C},\mathbf{R})$ between code changes $\mathbf{\mathbf{C}}$ and reviews $\mathbf{\mathbf{R}}$. 

\begin{equation}
   \mathbf{a}_{C,R} = [\mathbf{a}_{C}; \mathbf{a}_{R}],
\end{equation}
\begin{equation}
    Rel(C,R) = \tanh(\mathbf{w}^{\Lambda}\mathbf{a}_{C,R}),
\end{equation}

\noindent where $\mathbf{W}^{\Lambda}$ is the matrix of trainable parameters and $Rel(C,R)$ indicates the predicted relevancy score between one code change and review.

\subsection{Model Training and Testing}
\subsubsection{Training Setting} Since CORE aims at scoring the more related reviews higher given one code change, we determine the training goal as the \textit{Mean Square Error}~\cite{mse} loss function. 

\begin{equation}
    Loss = \frac{1}{N} \sum_{i=1}^{N} (Rel(c_i,r_i) - \hat{Rel}(c_i,r_i))^2,
\end{equation}
\noindent where $Rel(c_i,r_i)$ is the true relevancy label, $\hat{Rel}(c_i,r_i)$ is the predicted result of CORE, and $N$ is the total number of code change-review pairs. We use ADAM \cite{Adam} as our optimizer, with a learning rate of $1e-4$. The number of epochs is set to 50 and we use a dropout rate of 0.2. The word embedding size is set to 300 and the one-hot embedding size for characters is set to 60. The number of hidden states for Bi-LSTMs is set to 400 and the dimension of the attention layer is set to 100. For training the neural networks, we limit the vocabularies of the two sources to the top 50,000 tokens that are most frequently used in code changes and reviews. For implementation, we use PyTorch~\cite{Pytorch}, an open-source deep learning framework, which is widely-used in previous research~\cite{guo2019empirical,feng2019mobidroid}. We train our model in a server with one Tesla P40 GPU with 12GB memory. The training lasts ~35 hours.

\subsubsection{Testing Setting} We test our model using the GPU as above. Each testing phase took about 10 minutes. Each review is ranked with a random set of 50 reviews which only one of which is the review with the true label. 

\section{Experimental Setup}\label{sec:setup}
\subsection{Data Preparation}
We crawled the experimental datasets from GitHub, where the communities frequently submit pull requests to many open-source projects. We selected the projects for collection based on two criteria: The projects are i) popular JAVA projects on GitHub - to ensure the quality of the pull-request pairs; ii) projects with enough pull-request pairs - which necessitates an automated code review recommendation for code changes. To obtain projects that satisfy the two criteria, we randomly inspect the projects ranked at the top 200 on GitHub in terms of the number of stars, and keep the ones with more than 400 pull requests. We selected 19 projects, including Spark~\cite{ApacheSpark}, Neo4j~\cite{neo4j}, and elasticsearch~\cite{elasticsearch}. For each selected project, we created a GitHub API crawler to collect code changes and corresponding reviews. We ran our crawler in July 2019.
In total, we crawled 85,423 reviews from the 19 projects.

To further ensure the quality of the experimental datasets, we conducted preprocessing. We first filter out the reviews which are acknowledgement or feedback from the pull request author. 
This is because their replies are commonly acknowledgement or discussion of any feedback from the reviewer. Then we eliminate the reviews that are not written in English, and convert all the remained reviews into lowercase. We also conduct word lemmatization using NLTK\cite{nltk}, where each word is converted into its base or dictionary form. After removing empty review texts, we finally obtained 57,260 \textless code change, review\textgreater\, pairs. We randomly split the dataset
by $7:0.5:2.5$, as the training, validation, and test sets, \textit{i.e.}, there are 40,082, 2,863, and 14,315 pairs in the training, validation, and test sets, respectively.

We label the relevancy scores of all the 57,260 change-review pairs as 1, \textit{i.e.}, these pairs are regarded as ground truth or positive samples. To ensure that the model can also learn irrelevant pairs, we generate negative samples. We follow the typical learning-based retrieval strategies\cite{NIPS2014_5550}. Specifically, we randomly select $m$ reviews corresponding to the other code changes as negative reviews of the current code change, \textit{i.e.},  \textless$c_i, r_j$\textgreater \, where $i\neq j$ and the number of $r_j$ equals $m$. In this work, we experimentally set $m=5$. Detailed statistics of the experimental datasets are shown in Table~\ref{tbl-dataset}.

\begin{table}
  \centering
  \addtolength{\tabcolsep}{-3.5pt}
	\caption{Statistics of collected data}
	\label{tbl-dataset}
	\scalebox{0.9}{\begin{tabular}{l c c c }
	  \toprule 
    & \textbf{Training Data} & \textbf{Validation Data} & \textbf{Testing Data} \\
    \midrule\midrule
     \#Positive Samples & 40,082 & 2,863 & 14,315\\
    \midrule
     \#Negative Samples & 200,410 & 14,315 & 71,575\\
    \midrule
    Total & 240,492 & 17,178 & 85,890 \\
    \bottomrule
	\end{tabular}
	}
	
\end{table}

\subsection{Evaluation Metrics}  \label{ssec:metrics}
We adopt two common metrics for validating the effectiveness of code review recommendation, namely, $Recall@k$ \cite{DBLP:journals/corr/abs-1206-4603, DBLP:journals/corr/abs-1209-6492} and Mean Rank Reciprocal (MRR) \cite{7372014, inproceedings}, which are widely used in information retrieval and the code review generation literature \cite{radev-etal-2002-evaluating, trec-8, Severyn:2015:LRS:2766462.2767738}.

$Recall@k$ measures the percentage of code changes for which more than one correct result could exist in the top $k$ ranked results \cite{DBLP:journals/corr/abs-1206-4603, DBLP:journals/corr/abs-1209-6492} , calculated as follows:

\begin{equation}
    Recall@k = \frac{1}{|C|} \sum_{c\in C} \delta(Rank_c\leq k),
\end{equation}

\noindent where $C$ is a set of code changes, $\delta(\cdot)$ is a function which returns 1 if the input is true and 0 otherwise, and $Rank_c$ is the rank of correct results in the retrieved top $k$ results. Following prior studies \cite{Gupta2018IntelligentCR}, we evaluate $Recall@k$ when the value of $k$ is 1, 3, 5, and 10. $Recall_k$ is important because a better code review recommendation engine should allow developers to discover the needed review by inspecting fewer returned results. The higher the metric value is, the better the performance of code review recommendation is.

Mean Reciprocal Rank (MRR) is the average of the reciprocal ranks of results for a set of code changes $C$. The reciprocal rank of a code change is the inverse of the rank of the first hit result\cite{7372014, inproceedings}. MRR is defined as follows:

\begin{equation}
    MRR = \frac{1}{|C|} \sum_{c\in C} \frac{1}{Rank_c}.
\end{equation}

\noindent The higher the metric value is, the better the performance of code review recommendation is.

\subsection{Baselines for Comparison}
We compare the effectiveness of our approach with \textit{TF-IDF+LR} (Logistic Regression with Term Frequency–Inverse Document Frequency)\cite{Gupta2018IntelligentCR, tf-idf, 8226700} and a deep-learning-based approach, \textit{DeepMem}~\cite{Gupta2018IntelligentCR}.

\textit{TF-IDF+LR} is a popular, conventional text retrieval engine. It computes the relevancy score between one code change and review text based on their TF-IDF (Term Frequency–Inverse Document Frequency)\cite{SALTON1988513} representations. The training process is implemented by using the logistic regression (LR) method\cite{8226700}. Specifically, we concatenate the TF-IDF representations of code changes and reviews as the input of the LR method. 

\textit{DeepMem} \cite{Gupta2018IntelligentCR} is a state-of-the-art code review recommendation engine proposed recently {by Microsoft}. It recommends code reviews based on existing code reviews and their relevancy with the code changes. To learn the relevancy between the review and code changes, the review, code changes and the context (\textit{i.e.} three statements before and after the code changes) are input into a deep learning network for learning. They employed the use of LSTM for learning the relevancy between code changes and reviews. We use similar settings according to their paper, such as LSTM dimensions and model components. Since the dataset is not publicly released by the authors~\cite{Gupta2018IntelligentCR} and our crawled data do not contain context information of code changes, we only take code changes and reviews as the input of DeepMem.

\section{Evaluation} \label{sec:evaluation}
According the experimental setup in the above section, we conduct a quantitative analysis to evaluate the effectiveness of CORE in this section. In particular, we aim at answering the following research questions.

\begin{itemize}
\item \textbf{RQ1}: What is the accuracy of CORE?
\item \textbf{RQ2}: What is the impact of different modules on the performance of CORE? The modules include word-level embedding, character-level embedding, and the attention mechanism.

\item \textbf{RQ3}: How accurate is CORE under different parameter settings?

\end{itemize}

\begin{table}[t]
	\centering

	\caption{Comparison results with baseline models. $R@K$ indicates the metric $Recall@K$. Statistical significance results are indicated with * for $p-value<0.01$.}
	\label{tbl-evaluations}
	\scalebox{0.9}{\begin{tabular}{l c c c c c}
		\toprule 
		\textbf{Model} & \textbf{MRR} & \textbf{R@1} & \textbf{R@3} & \textbf{R@5} & \textbf{R@10} \\
		\midrule\midrule
		\textbf{TF-IDF+LR} & 0.089* & 0.019* & 0.060* & 0.100* & 0.201* \\
		\midrule
		\textbf{DeepMem} & 0.093* & 0.021* & 0.065* & 0.108* & 0.208*\\
		\midrule
		\textbf{CORE} & \textbf{0.234} & \textbf{0.113} & \textbf{0.247} & \textbf{0.333} & \textbf{0.482}\\
		\bottomrule
	\end{tabular}
	}

\end{table}

\subsection{RQ1: What is the accuracy of CORE?}
The comparison results with the baseline approaches are shown in Table~\ref{tbl-evaluations}. We can see that our CORE model outperforms all baselines. Specifically, the result of the non-deep-learning-based TF-IDF+LR model achieves the lowest performance, with 0.019 in Recall@1, 0.201 in Recall@10, and 0.089 in MRR score. The result is consistent with the finding by Gupta and Sundaresan~\cite{Gupta2018IntelligentCR}. This indicates that deep-learning-based models tend to better learn the semantic consistency between code changes and reviews.

CORE can increase the performance of DeepMem by 438.1\% in Recall@1, 131.0\% in Recall@10, and 150.7\% in the MRR score. We then use t-test and effect size measures for statistical significance test and Cliff's Delta (or $d$) to measure the effect size~\cite{DBLP:journals/technometrics/Ahmed06}. The significance test result ($p-value<0.01$) and large effect size ($d>1$)  on all the five metrics of CORE and DeepMem/TF-IDF+LR confirm the superiority of CORE over TF-IDF+LR {and DeepMem}. This explains that the reviews recommended by CORE are more relevant to the code changes than those from {TF-IDF+LR and} DeepMem.

\subsection{RQ2: What is the impact of different modules on the performance of CORE?}
We study the impact of each of the three modules, including character-level embedding, word-level embedding, and the multi-attention network, on the performance of CORE. We evaluate the model when each of the modules is removed individually, \textit{i.e.}, only one module is removed from CORE in each experiment. The comparison results are listed in Table~\ref{tbl-ablation}.

As can be observed in Table~\ref{tbl-ablation}, the combination of all modules achieve the highest improvements in performance. With only word-level embedding or character-level embedding involved, the model can only achieve comparable results as DeepMem. This indicates that the two-level embeddings can compensate for each other for better representing code changes and reviews. By comparing CORE with CORE without the multi-attention network, we can find that the strength brought by the attention mechanism is not that obvious. Without considering attention, the model can achieve slightly higher performance than that with attention in terms of Recall@10 and Recall@5. With attention involved, CORE returns better results for Recall@1. Since developers generally prefer relevant reviews to be ranked higher, our proposed CORE can be regarded as more effective.

\begin{table}[t]\footnotesize
	\centering
	\caption{Comparison results with different module removed. The ``CORE-WV'', ``CORE-CV'', and ``CORE-ATTEN'' indicate the proposed CORE without considering word-level embedding, character-level embedding, the multi-attention network, respectively.
	}
	\label{tbl-ablation}
	\scalebox{0.9}{\begin{tabular}{l c c c c c}
		\toprule 
		\textbf{Model} & \textbf{MRR} & \textbf{R@1} & \textbf{R@3} & \textbf{R@5} & \textbf{R@10} \\
		\midrule\midrule
		\textbf{CORE-WV} & 0.091 & 0.020 & 0.062 & 0.102 & 0.202\\
		\midrule
		\textbf{CORE-CV} & 0.103 & 0.026 & 0.077 & 0.123 & 0.233\\
		\midrule
		\textbf{CORE-ATTEN} & 0.233 & 0.107 & 0.246 & \textbf{0.339} & \textbf{0.498}\\
		\midrule
    	\textbf{CORE} & \textbf{0.234} & \textbf{0.113} & \textbf{0.247} & 0.333 & 0.482\\
		\bottomrule
	\end{tabular}
	}
	\vspace{-2mm}
\end{table}

\subsection{RQ3: How accurate is CORE under different parameter settings?}

We compare the performance of CORE under different parameter settings. We conduct the parameter analysis for the number of negative samples for each positive code change-review pair, the number of hidden units, and embedding size, respectively. We vary the values of these three parameters and evaluate their impact on the performance.

\begin{figure}[h]
    \centering
	\subfigure[Impact of \#negative samples on recall rate]
	{\label{fig:a}\includegraphics[width=0.28\textwidth]{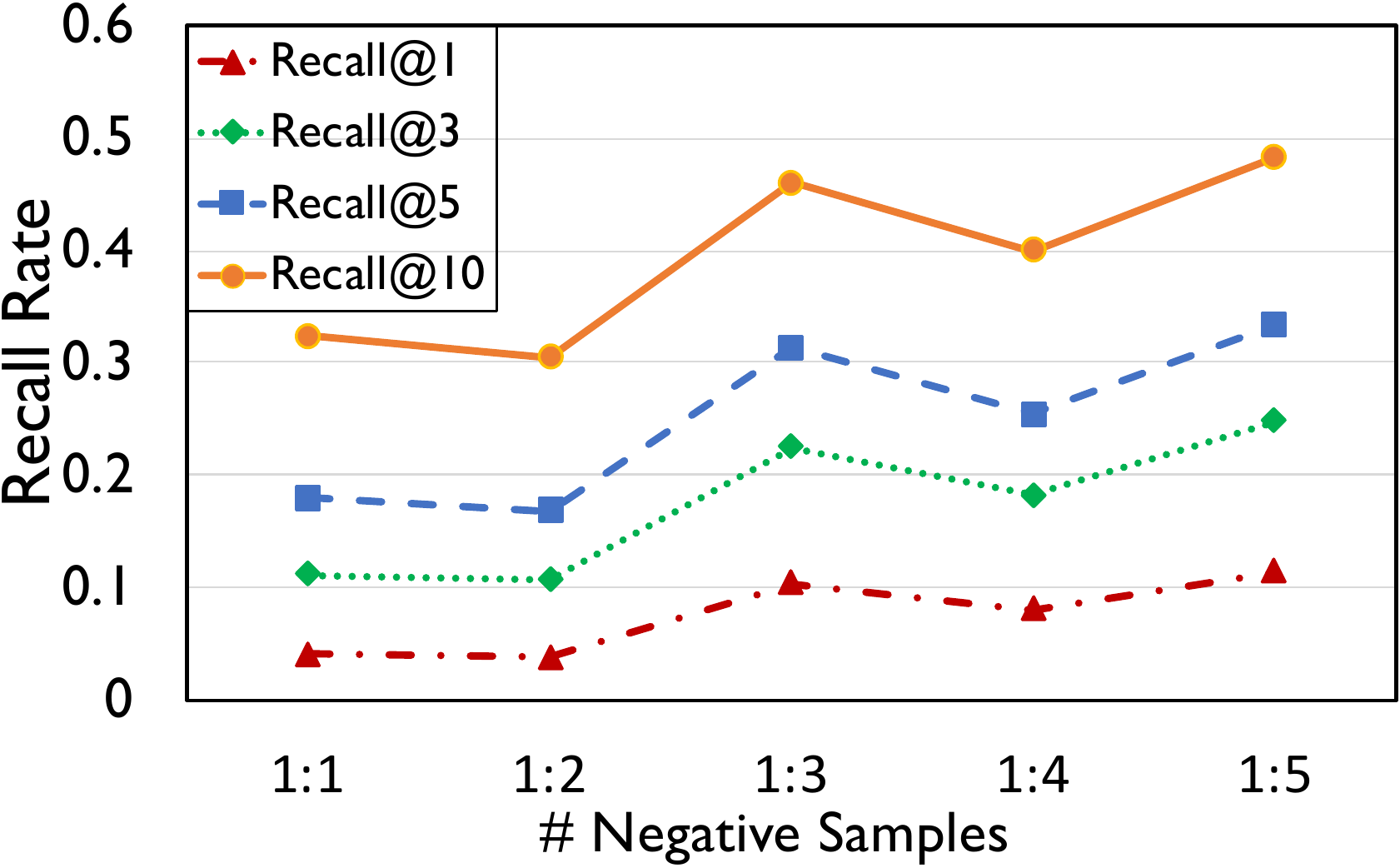}}
	\subfigure[Impact of \#hidden units on recall rate]
	{\label{fig:b}\includegraphics[width=0.28\textwidth]{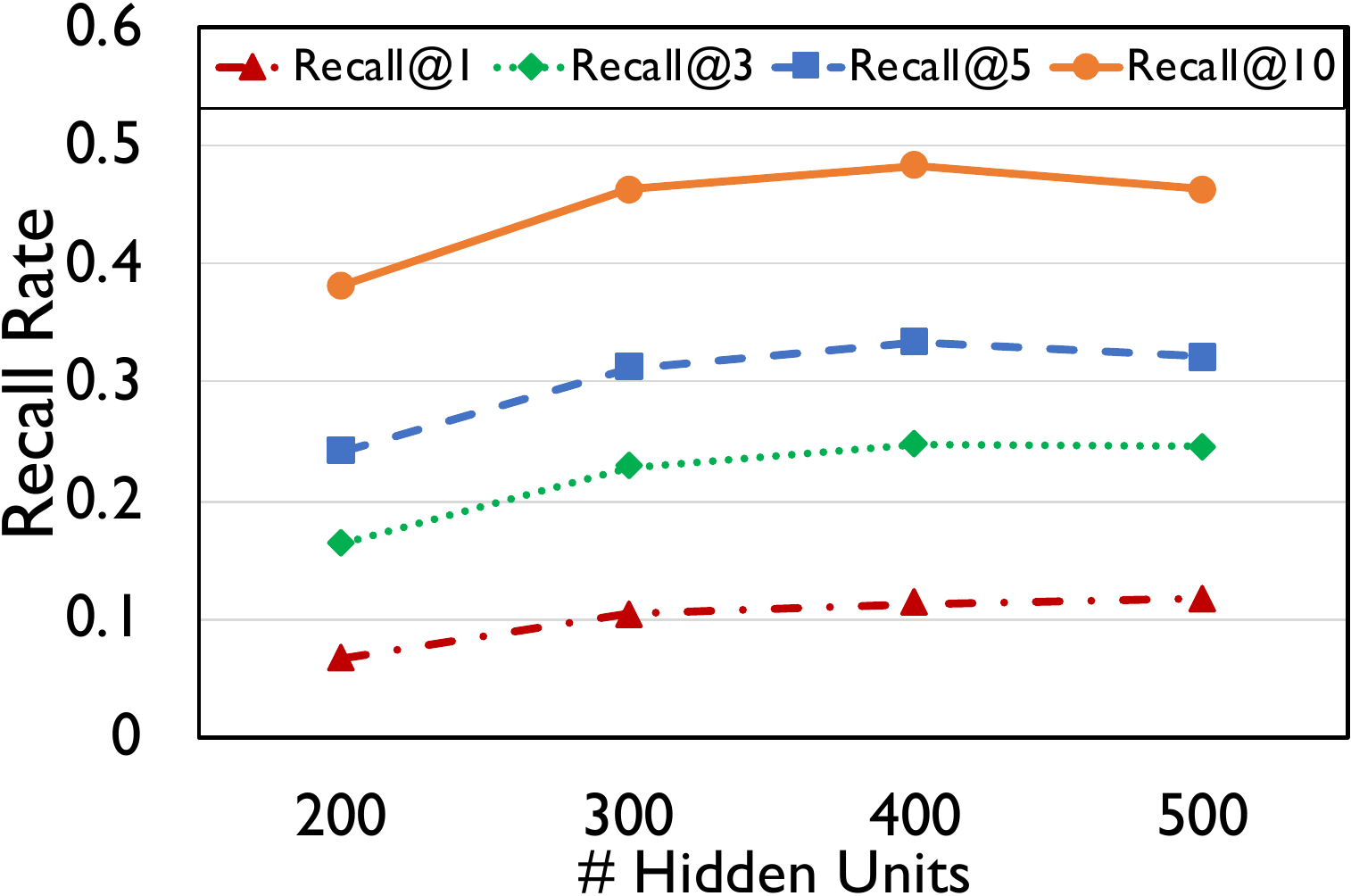}}
	
	\subfigure[MRR with \#negative samples]
	{\label{fig:c}\includegraphics[width=0.18\textwidth]{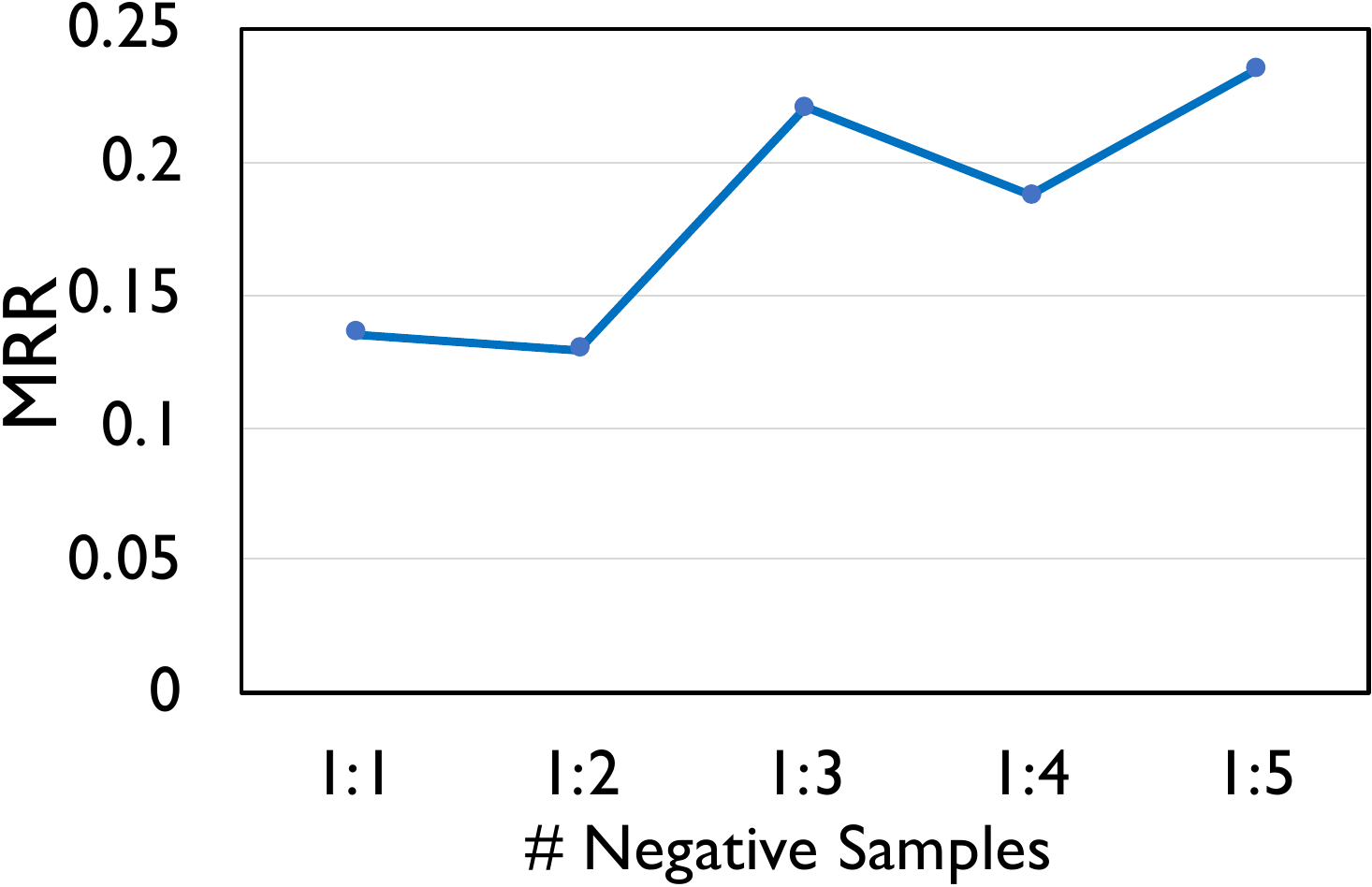}}
	\subfigure[MRR with \#hidden units]
	{\label{fig:d}\includegraphics[width=0.18\textwidth]{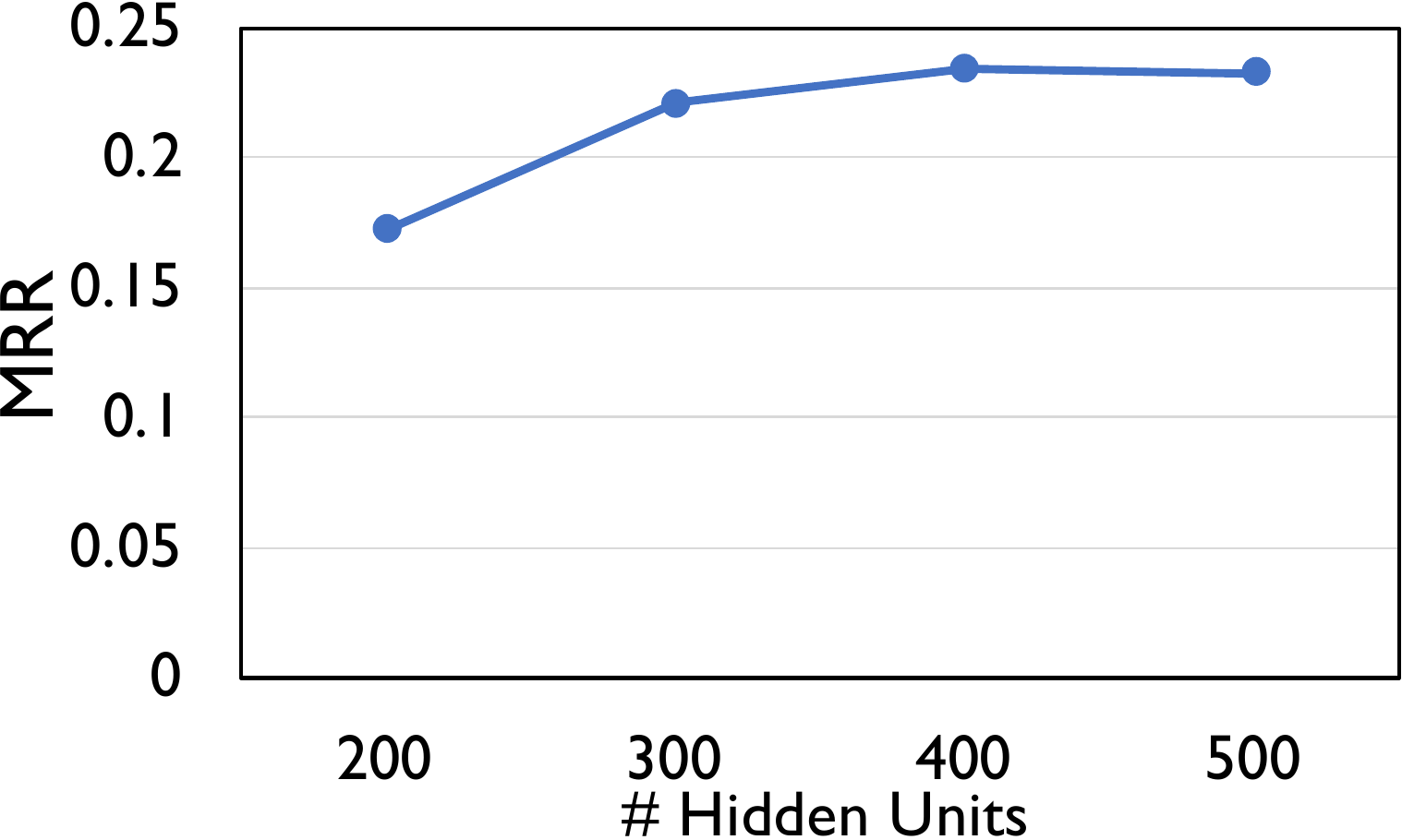}}
	
	\subfigure[Impact of different dimensions of word-level embedding]
	{\label{fig:e}\includegraphics[width=0.35\textwidth]{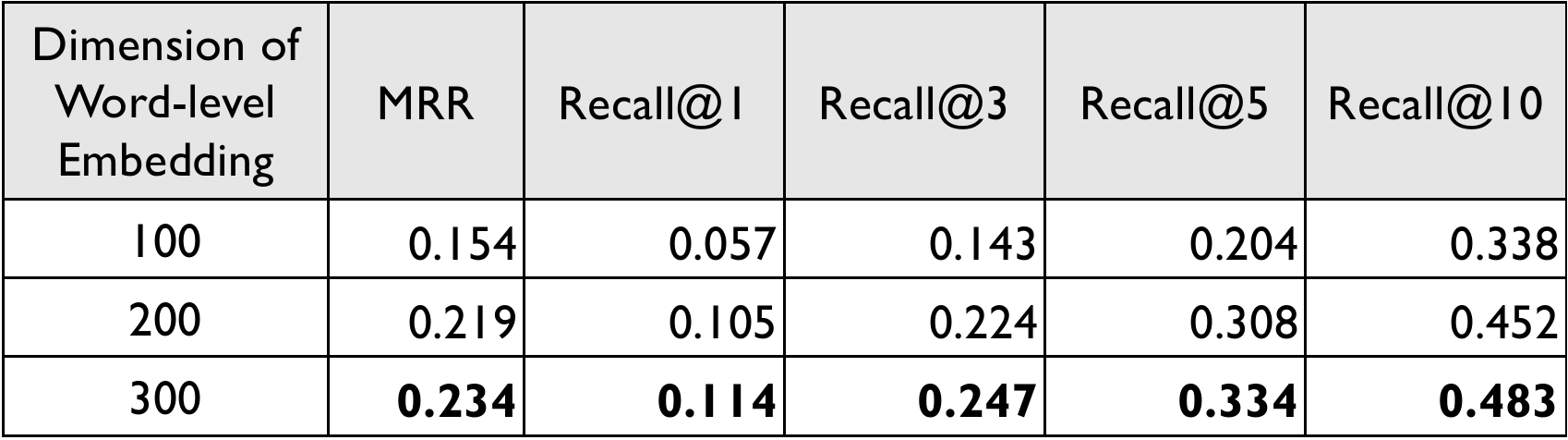}}
   
	 \caption{Performance under different parameter settings}

    \label{fig:para}
\end{figure}

Fig.~\ref{fig:a} and Fig.~\ref{fig:c} show the performance of CORE when different numbers $m$ of negative samples are randomly selected for each collected \textless code, review\textgreater\, pair. We range $m$ from 1 to 5, denoted as 1:1, 1:2, 1:3, 1:4, and 1:5. As can be seen, the performance of CORE shows a general upward trend in spite that the trend is non-monotonic. Since more negative samples can increase the time cost for model training, we set $m=5$ for balancing the time cost and performance.

According to Fig.~\ref{fig:b} and Fig.~\ref{fig:d}, the performance of CORE also varies along with different numbers of hidden units. We can see that more hidden units may not help improve accuracy. CORE generates the best result when we define the number of hidden units as 400.
Fig.~\ref{fig:e} shows the performance of CORE under different dimensions of word-level embeddings. As can be seen, more dimensions can benefit CORE in terms of all metrics. This explains that higher dimensions of word-level embedding can help better learn the representations of code changes and reviews. We can observe that the increase in performance slows down with the embedding dimension growing in 100. We set the dimension of word-level embedding as 300 due to its optimum performance.

\section{User Study}
In this section, we conduct a human evaluation to further verify the effectiveness of CORE in automating code review. As the effectiveness evaluation may be subjective, we consulted {industrial} developers to assess the effectiveness of CORE. {We collaborated with {Alibaba},

which is one of the largest e-commerce companies worldwide and the developers are proficient in software development. We set up several online meetings with them and conduct our human evaluation in the company with their help. Our human evaluation involves 12 front-line Java developers and code reviewers that have at least 3 years of coding and code reviewing experience. Specifically,}
we applied CORE to the 19 popular Java projects hosted on GitHub, and randomly selected 10 recommended reviews for code changes from each project. We asked the two questions shown in Table~\ref{tab:survey} to developers.

Developers showed great interest in CORE, and all the developers agreed that our tool can help them in practical development. For the answers to the second question, the developers point out the useful aspects and the aspects that need further improvement. Specifically, the useful aspects given by the developers include relevant reviews that CORE recommended and suggested actions in the recommended review. The aspects for enhancement are mainly about the recommendation of the code changes for review (\textit{e.g.}, pointing to the second line and giving review ``\textit{There exists a potential NullPointerException issue}''.) and highlighting the review keywords. 

In fact, for the industrial companies, it is a great demand for project development and maintenance to automatically generate code review. The code review process is a critical phase to ensure the code quality of the project, however, it costs substantial human effort. The demand for automated code review generation will be more urgent, especially for companies like {Alibaba} which provide services to a large number of users.

\begin{table}\footnotesize
\centering
\caption{Questions in the developer survey}
	\label{tab:survey}
\begin{tabular}{@{}lc@{}}
\toprule
Questions & \#Participants \\ \midrule
\multirow{3}{*}{\begin{tabular}[c]{@{}l@{}}Q1. Do you think the recommended reviews are\\ effective for practical development?\end{tabular}} & \multirow{3}{*}{12} \\
 &  \\
 &  \\ \midrule
\multirow{3}{*}{\begin{tabular}[c]{@{}l@{}}Q2. If you think the automated code review tool \\ is useful, which parts make you think so? If not \\ useful, which parts?\end{tabular}} & \multirow{3}{*}{12} \\
 &  \\
 &  \\ \bottomrule
\end{tabular}

\end{table}

\section{Discussion} \label{sec:discussion}
In this section, we discuss the advantages and limitations of CORE. We also illustrate the threats of validity of our work regarding the dataset and evaluation.

\subsection{Why does CORE work?} \label{subsec:core-working}
We have identified two advantages of CORE that may explain its effectiveness in code review recommendation.

\vspace{1mm}
\noindent\textbf{Advantage 1: CORE can well learn representations of source code and reviews.} 
CORE represents the code changes and reviews based on the two-level (including character-level and word-level) embeddings and also multi-attention network. The two-level embeddings can well capture the semantics of both code changes and reviews, and the multi-attention network allows CORE to focus on the words and characters that are representative of code changes and reviews. In this way, CORE can well learn the representations of both sources. We highlight two real cases to demonstrate the advantages.

\begin{lstlisting}[language=Java,caption={Case study on retrieving top review (return message)},label={fig-cs-return-msg}]
// Code changes
-  return "[" + nodeId + "][" + taskId + "] failed, reason [" + getReason() + "]";
------------------------------------------------
// Review by CORE
    "please add a message otherwise this just throws a nullpointerexception with no helpful message about what was null"
// Review by DeepMem
    "do we need to fully qualify this io netty handler codec http2 http2stream state http2stream state"
// Review by TF-IDF+LR
    "do we need to fully qualify this io netty handler codec http2 http2stream state http2stream state"
\end{lstlisting}

As observed in Listing \ref{fig-cs-return-msg}, CORE recommends the most relevant review for the code change, providing actions such as adding a message or warning about an exception.
CORE can learn the relevancy between semantically-related tokens, such as ``\textit{failed}" in the code and ``\textit{NullPointerException}" in the review. The two tokens are generally used to express an exception or error that occurs unexpectedly. But both reviews recommended by the baselines do not capture the failure-related token in the code changes and produce wrong results. We also discover that DeepMem and TF-IDF+LR prioritize the same review, which may be due to both models focus on general words, such as "\textit{return}" and "\textit{nodeID}"

\begin{lstlisting}[language=Java,caption={Case study on retrieving top review (test cases)},label={fig-cs-test-cases}]
// Code changes
-package org.elasticsearch.node;
-import org.elasticsearch.common.collect.ImmutableOpenMap;
-import org.elasticsearch.common.settings.Settings;
-import org.elasticsearch.test.ESTestCase;
-import static org.hamcrest.Matchers.equalTo;
-public class NodeModuleTest extends ESTestCase{
- public void testIsNodeIngestEnabledSettings(){
------------------------------------------------
// Review by CORE
    "can you split this into two different tests rather than using the randomness here i dont think it buys us anything"
// Review by DeepMem
    "i'd make the string here junk or not interpreted or something if you dont read the file carefully it looks like the script is run because that is a valid looking script"
// Review by TF-IDF+LR
    "we need the version check here as well for bw comp"
\end{lstlisting}

A similar case can be found in Listing~\ref{fig-cs-test-cases}. The code change is mainly about adding a new test case which can be well captured by CORE (\textit{e.g.}, the token ``\textit{tests}'' in the review can be well-matched with ``\textit{test}'' in the code). Such a semantic match cannot be observed in the recommended reviews of baselines. For example, the prioritized review of DeepMem discusses string manipulation and the top review of TF-IDF+LR talks about versioning.

\vspace{1mm}
\noindent\textbf{Advantage 2: CORE can better solve Out-of-Vocabulary (OOV) issues.} 
To solve the OOV issue in code changes, CORE builds upon character-level embedding. We present two real cases for illustrating the effectiveness of character-level embedding on the OOV issue. 

\begin{lstlisting}[language=Java,caption={Case study on comparing CORE-WV \& CORE},label={fig-cs-core-wv}]
// Code changes
- deleteSnapshot(snapshot.snapshotId(), new DeleteSnapshotListener(){
+ deleteSnapshot(snapshot.snapshotId().getSnapshot(), new DeleteSnapshotListener() {
------------------------------------------------
// Review by CORE
    "can you add a comment here as to why we have a special handling for external versions and deletes here"
// Review by CORE-WV
    "should values lower than versionnum be allowed here"
\end{lstlisting}

\begin{lstlisting}[language=Java,caption={Case study on comparing CORE-WV \& CORE 2},label={fig-cs-core-wv-2}]
// Code changes
-  inSyncAllocationIds.contains(shardRouting.allocationId().getId()) == false)
+  inSyncAllocationIds.contains(shardRouting.allocationId().getId()) == false &&
+  (inSyncAllocationIds.contains(RecoverySource.ExistingStoreRecoverySource.FORCED_ALLOCATION_ID) == false
------------------------------------------------
// Review by CORE
    "can we remove this allocation"
// Review by CORE-WV
    "that s just a minor thing but i think the recommended order in the java styleguide is static final"
\end{lstlisting}

As observed in Listing~\ref{fig-cs-core-wv}, CORE recommends a review suggesting to add comments for the handling and the deletes. The review is strongly relevant to the code change which revolves around deletion. Without the character-level embedding considered, CORE-WV ranks a different and unrelated review at the top. One possible reason is that CORE-WV could not well learn the semantic representation of ``\textit{deleteSnapshotListener}'' and ``\textit{deleteSnapshot}'' due to the low co-occurrence frequency\footnote{The token ``\textit{deleteSnapshotListener}'' only appears 349 times among the whole GitHub repository~\cite{frequency}.} of the subwords (such as ``\textit{delete}'' and ``\textit{snapshot}''). A similar case can be found in Listing~\ref{fig-cs-core-wv-2}, CORE-WV cannot well capture the semantics of the variable ``\textit{inSyncAllocationIds}'' while CORE can learn that the token is semantically related to ``\textit{allocation}''. Thus, with character-level embeddings, CORE can better focus on the important characters in the variable and function name.

\subsection{Limitations of CORE}
We show the limitations of CORE by using Listing \ref{fig-cs-3}. CORE ranks this example at position 30. One possible reason for the low relevancy score could be the lack of related keywords between the code change and the review. Without relevant keywords, (\textit{e.g.}, shown in Section \ref{subsec:core-working}), CORE fails to capture the relevancy between the two of them. Furthermore, the review contains only generic keywords (\textit{e.g.}, ``\textit{supposed}", ``\textit{were}") and does not have any keywords relevant to ``\textit{asset}" or ``\textit{HashSet}".
Despite that CORE does not retrieve the ground truth, its recommended review shows that CORE captures the main semantics of the code change. As can be seen in the code change, the function ``\textit{assetEquals}" is invoked to evaluate a condition. CORE understands that the code change is related to checks and assertions; hence, the retrieved review is relevant to an assertion. This further shows that CORE can focus on the key information in the code changes. {Moreover, CORE is flexible and extensible to integrate external information, and can involve source code (such as code structure) and code comments to better learn the semantic relevancy between code changes and reviews in future.}

\begin{lstlisting}[language=Java,caption={Case study on limitation of CORE},label={fig-cs-3}]
// Code changes
-  assertEquals(3, exec("def x = new HashSet(); x.add(2); x.add(3); x.add(-2); def y = x.iterator(); " +
-  "def total = 0; while (y.hasNext()) total += (int)y.next(); return total;"));
------------------------------------------------
// Ground Truth
    "were these supposed to be removed"
// Review by CORE
    "that check should be reversed we assert that it s not null the else part dealing with the case when it is"
\end{lstlisting}

We further investigate the relation between the length of code tokens and the performance of CORE. Table \ref{tbl-code-length} shows the MRR and Recall@K for different code lengths. Our experiments show that CORE performs better for the code changes with lengths ranging from 50 to 75. One possible reason might be that longer code sequences are trimmed to a fixed length, and only the beginning part of the sequence may not fully represent the semantics of the code changes.

\begin{table}[t]
	\centering
	\caption{Performance of CORE regarding different code token length}
	\label{tbl-code-length}
	\begin{tabular}{l c c c c c}
		\toprule 
		\textbf{Code Length ($l$)} & \textbf{MRR} & \textbf{R@1} & \textbf{R@3} & \textbf{R@5} & \textbf{R@10} \\
		\midrule\midrule
		\textbf{$l<$25} & 0.1954 &  0.0789 & 0.2171 & 0.2828 & 0.4342 \\
		\midrule
		\textbf{25$\leq l<$50} & 0.2186 & 0.1029 & 0.2242 & 0.3010 & 0.4685\\
		\midrule
		\textbf{50$\leq l<$75} & \textbf{0.2423} & \textbf{0.1246} & \textbf{0.2468} & \textbf{0.3441} & \textbf{0.5000}\\
		\midrule
		\textbf{$l>$75} & 0.2352 & 0.1138 & 0.2490 & 0.3353 & 0.4829\\
		\bottomrule
	\end{tabular}
	\vspace{-3mm}
\end{table}

\subsection{Threats to Validity}
\subsubsection{Subject Dataset}
One of the threats is the quality of code changes and reviews in the dataset. Overlap of data in the training and testing set has been a great issue among deep learning. One of the biggest concerns for our work is that for any pair of \textless code changes, review\textgreater, the negative data might exist in training set while one true positive data exists in the testing set, vice versa. This results in the model learning some parts of the testing set during the training phase. In our work, we ensure that our training and testing set do not have any overlapping pairs of \textless code changes, review\textgreater\, by splitting the data into two sets before we negatively sample them. Therefore, the training set will have its own set of positive and negative pairs while the testing set has its own set of positive and negative pairs.

\vspace{2mm}
\subsubsection{Comparison with DeepMem}
Another threat is that the results of the DeepMem in our implementation could be lower than the original model. The result on DeepMem in the original paper \cite{Gupta2018IntelligentCR} shows a Recall@10 of 0.227 and 0.2 MRR score. However, in our implementation of DeepMem, it only achieved 0.208 in Recall@10 and 0.1 in MRR score. This may be due to the reason that their method uses additional code contexts, \textit{i.e.}, three statements of code between and after the code changes, for additional context. We do not consider the context as their data are not publicly available and our crawled data do not contain such information. To combat such a threat, we carefully review the technical part of DeepMem in the published paper and confirm the implementation with the other three co-authors. Furthermore, we evaluate both CORE and the implemented DeepMem with similar settings and on the same dataset for a fair comparison.

\vspace{2mm}
\subsubsection{User Study}
The quality of user study might be a threat to the validity of this paper. Our user study involves some perspectives from the developers and the feedback might varies from each developers. We mitigate the threat by seeking developers with at least 3-5 years of coding experiences. Furthermore, we ensure that they have at least have prior experiences to code reviewing in the industry.

\section{Related Work}
\subsection{Automating Code Review}

The techniques for automating code review can be generally divided into static analysis and deep learning.

Static Analysis tools can provide a fast and efficient preliminary analysis of the source code. A code collaboration tool, ReviewBot \cite{6606642}, uses the results of existing static analysis tools and generates reviews and reviewer recommendations. A machine learning model, by Michal et al.\cite{8104731}, uses metrics such as information about the author, files attributes and source code metrics, to classify the code changes. Going beyond the machine learning and static analysis, Pavol et al. \cite{DBLP:journals/corr/BielikRV16} uses the results of static analysis tools and synthesis algorithms to learn edge cases that common static analysis tools could not find. This approach solved the overfitting problem that is commonly found in such methods.
There are several works that employ the use of deep learning and code review \cite{Gupta2018IntelligentCR}. One of the most relevant work, a model by Anshul Gupta et al. \cite{Gupta2018IntelligentCR}, uses information retrieval techniques, such as LSTM and fully connected layer to learn the relationship between source code and the reviews. A recent work, by Toufique et al. \cite{8115623}, uses Sentiment Analysis on code review to determine if the reviews are positive, neutral or negative comments. Shi et al.\cite{Shi2019AutomaticCR} presents a deep learning-based model that uses source codes that are before modification and after modification. The model could determine if the submitted code changes are likely to be approved or rejected by the project administrators. Chen et al.~\cite{chen2019storydroid} proposed to extract components like UI pages from Android apps to help the review process. Unfortunately, some of these works do not focus on review generation or retrieval. In those works that concern with code reviews retrieval or generation, they often require much more code contexts than just the code changes. 

\subsection{Code to Embedding}
Due to the nature of source code, several works focus on source code embedding. The most commonly used method in embedding source is by using word2vec \cite{Mikolov:2013:DRW:2999792.2999959}. Word2Vec models the distributions of the word in a large corpus and embed words into a common latent dimension. It is one of the most commonly used embeddings in both natural languages and source code. A work by Gu et al. \cite{8453172} uses several different features of the function to embed the source code. It uses method name, API sequences and tokens in the embedding function to obtain the word/token vector. Graph-based embedding, such as Code2Seq \cite{DBLP:journals/corr/abs-1808-01400}, uses Abstract Syntax Tree (AST) to embed code sequences to represent their underlying structure. These embedding techniques allow us to understand the inner representation of code structures, enhancing any code-related deep learning task. for instance, our work.  

\subsection{Code Summarization}
Summarization of source code has been a research problem for a long time. Despite the granularity of code summarization is much bigger, some similarities exist between code reviews and summarization. Several automated tools, such as Javadocs \cite{Kramer1999APIDF} and Doxygen \cite{doxygen}, can be used to provide comments for source code. Gu et al. \cite{8453172} uses multiple features to embed source codes and uses them to search for similar comments. Iyer et al. \cite{bba651a5f9344ec2b3c17a463eeeb077} uses LSTM with attention to perform code summarization. Hu et al. \cite{Hu:2018:DCC:3196321.3196334} uses Seq2Seq and AST embedding to provide a more sophisticated deep learning approach for the same problem. Much earlier works, such as Haiduc \cite{6062165}, implemented summarization by text mining and text retrieval methods. Some papers \cite{Sridhara:2010:TAG:1858996.1859006,Rodeghero2015AnES} also researched on heuristic-based and natural language techniques to provide comments for source code of small functions/methods.

\section{Conclusion}
We proposed a novel multi-level embedding attentional neural network, CORE, for learning the relevancy between code changes and reviews. Our model is based on word-level and character-level that aims to capture both the semantic information in both source code and reviews. In the future, the generation of reviews could be improved by using neural translation and better embedding for code changes. 

\section*{Acknowledgement}
We appreciate the constructive comments from reviewers. This work is supported by Alibaba Cloud Singapore Grant AN-GC-2018-019. The authors would like to thank Nvidia for their GPU support.

\clearpage
\balance
\bibliographystyle{IEEEtran}
\bibliography{ref}

\end{document}